\documentclass[10pt,twocolumn,twoside]{IEEEtran}
\IEEEoverridecommandlockouts

\usepackage{cite}
\usepackage{amsmath,amssymb,amsfonts}
\usepackage{algorithmic}
\usepackage{graphicx}
\usepackage{textcomp}
\usepackage{balance}
\usepackage{xcolor}
\usepackage[utf8]{inputenc}
\usepackage{tikz}
\usepackage{graphicx}
\usepackage{adjustbox}
\usepackage{booktabs}
\usepackage{multirow}
\usepackage{float}
\usepackage{url}

\def\BibTeX{{\rm B\kern-.05em{\sc i\kern-.025em b}\kern-.08em
    T\kern-.1667em\lower.7ex\hbox{E}\kern-.125emX}}

\title{\LARGE \bf GPU-to-Grid: Voltage Regulation via GPU Utilization Control}
\author{Zhirui Liang, Jae-Won Chung, Mosharaf Chowdhury, Jiasi Chen, Vladimir Dvorkin
\thanks{All authors are with the University of Michigan, Ann Arbor, MI, USA.}%
}

\begin{document}
\begingroup
\allowdisplaybreaks

\maketitle

\begin{abstract}

While the rapid expansion of data centers poses challenges for power grids, it also offers new opportunities as flexible loads. Existing power system research often abstracts data centers as aggregate resources, while computer system research focuses on GPU energy efficiency and largely ignores grid impacts. To bridge this gap, we develop a GPU-to-Grid framework that couples device-level GPU control with power system objectives. We study distribution-level voltage regulation enabled by LLM inference flexibility, using batch size as a data-center-side control knob that trades off GPU power consumption, inference latency, and token throughput. We first formulate the problem as an optimization problem and then realize it as an online feedback optimization controller, implemented by the data center operator using its own empirical GPU power-performance model and real-time measurements from both the GPU and grid systems. Our key insight is that reducing GPU power alleviates lower-voltage violations, while increasing GPU power mitigates upper-voltage violations; this challenges the common belief that minimizing GPU power is always beneficial to power grids.
\footnote{Open-sourced as part of the OpenG2G library~\cite{openg2g-arxiv26,openg2g-github}.}
\end{abstract}



\section{Introduction} \label{sec:introduction}

The rapid expansion of AI workloads is driving a sharp rise in data center electricity demand and GPU power density. Globally, data centers consumed about 415~TWh in 2024 and are projected to more than double by 2030, with AI as a key growth driver \cite{LBNL2024DataCenterEnergy}. At the device level, state-of-the-art accelerators already approach server-scale power intensities (e.g., NVIDIA’s H100 SXM5 GPU specifies up to 700~W per GPU \cite{NVIDIAH100}), so large GPU clusters can add multi-megawatt loads over short deployment timelines. This surge creates significant challenges for power-system operation and planning: load growth is geographically concentrated, often constrained by latency and reliability requirements, and can substantially change regional demand trajectories \cite{ai-grid-impact-arxiv25}.

At the same time, grid-connected data centers also offer new opportunities for power system operation. Similar to vehicle-to-grid (V2G) technologies~\cite{guille2009conceptual}, large-scale data centers can act as flexible demand-side resources by adjusting power consumption in response to grid conditions.
Recent work in the power system literature has explored the use of data center flexibility for services such as peak shaving, frequency regulation, and voltage support \cite{dvorkin2024agent,fu2020assessments,xie2025enhancing}.
However, most existing studies model data center flexibility at an aggregate level and do not explicitly capture how such flexibility is realized at the device level. In particular, the mechanisms by which GPU workloads provide controllable power adjustments, along with the associated inference latency and throughput constraints, are often abstracted or neglected. These GPU-level considerations are critical in practice, as they enable fine-grained and fast power control while limiting the achievable magnitude and speed of power adjustments.

From the computer systems perspective, significant effort has been devoted to improving the energy efficiency of GPUs through workload-aware control. Both training and inference energy consumption can be reduced by tuning available control knobs, such as GPU frequency scaling, power caps, and batch size selection, which directly affect utilization, throughput, and latency \cite{you2023zeus,perseus-sosp24,mlenergy-neuripsdb25}. While these methods optimize energy or performance under fixed workload objectives, they do not account for grid needs. From the power system standpoint, there are operating conditions under which increased GPU power consumption is desirable, for example during periods of high renewable generation or when overvoltage arises in distribution networks. Enabling GPUs to act as grid-supportive resources therefore requires closing the loop between grid conditions and device-level control decisions. 

To bridge this gap, we propose a GPU-to-Grid (G2G) framework that couples device-level GPU control with grid-level feedback. The framework integrates models of GPU power-performance trade-offs with real-time grid signals, enabling GPUs to operate as grid-supportive resources while respecting computing constraints. Such grid feedback signals may take the form of voltage measurements, frequency deviations, or price signals, depending on the grid service being provided.

In this paper, we demonstrate the G2G framework using one grid service, distribution-level voltage regulation, and one GPU control knob, the batch size of LLM inference tasks. The controller is implemented by the data-center operator, which updates batch size locally in response to limited grid voltage and LLM latency measurements. Fig.~\ref{fig:architecture} shows the overall architecture. Users submit stochastic inference requests to heterogeneous LLM models served by dedicated GPUs. The resulting batch-size decisions affect user latency, token throughput, GPU power consumption, and distribution-network voltages. Thus, at each control interval, the controller balances voltage constraints, latency requirements, and data-center throughput objectives within the proposed G2G framework.

Several studies have investigated grid services using device-level models of data center resources. For instance, Chen et al. \cite{chen2025voltage} employ GPUs for voltage regulation via dynamic voltage and frequency scaling (DVFS), but assumes a linear relationship between GPU frequency and power and adopts a simple droop-based control that ignores inference latency and throughput constraints. In contrast, Colangelo et al. \cite{colangelo2025ai} demonstrate grid-interactive AI data centers in a field deployment, showing that workload control and DVFS can reduce power consumption during peak periods while maintaining quality of service; however, the grid feedback in that work is limited to high-level exogenous signals such as congestion or peak-demand indicators, rather than physical states of power systems. This work addresses these limitations by using real GPU measurement data to model the nonlinear relationships between control knobs and performance metrics, and by proposing an online-feedback-optimization framework that explicitly incorporates voltage, latency, and throughput. By closing the loop with physical grid measurements and avoiding reliance on demand-response signals, the proposed approach enables fast distribution voltage regulation.


\begin{figure}[t]
\centering
\resizebox{\columnwidth}{!}{%
\begin{tikzpicture}[
    circ_node/.style={circle,scale=0.5,color=black,fill=black},
    sub_station/.style={circle,scale=1.0,inner sep=5pt,draw=black,line width=0.025cm},
]

\node [draw=blue, line width=0.01cm, dashed, rounded corners=2.5, minimum width=190, minimum height=55, text depth=0em, align=left] at (0.25,0) {};
\node[draw=none,font=\small,blue] at (-1.6,1.2) {Power system simulator};
\node[sub_station] (n00) at (-2.25,0) {};
\node[sub_station] (n01) at (-2,0) {};
\node[below of = n01, yshift=0.3cm,  xshift=-0.125cm, align=center] {Substation};
\node[circ_node,label=90:{$v_{1}(p)$}] (n1) at (0,0) {};
\node[circ_node,label=90:{$v_{2}(p)$}] (n2) at (1.5,0) {};
\node[circ_node,label=90:{$v_{n}(p)$}] (n3) at (3,0) {};
\draw[thick,black] (n01) -- (n1) -- (n2) -- node[pos=0.5,fill=white] {$\dots$} (n3);

\node [draw=black, line width = 0.025cm, rounded corners = 2.5,minimum width=130, minimum height=25, text depth=0em, align=center] (controller) at (5,1.8) {$b\leftarrow b - \rho_b \nabla\mathcal{L}(\mathbf v(p(b)),l(b), r(b))$};

\node [draw=none, align=center] at (5,2.5) {batch size controller};

\node [draw=blue, line width=0.01cm, dashed, rounded corners=2.5, minimum width=193, minimum height=55, text depth=0em, align=left] at (2.5,-3.25) {};
\node[draw=none,font=\small,blue] at (3.2,-4.5) {Cluster simulator based on real measurements};
\node [draw=black, line width = 0.025cm, rounded corners = 2.5,minimum width=40, minimum height=40, text depth=0em, align=center] (gpu) at (0,-3.25) {GPUs \\ \includegraphics[width=1cm]{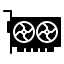}};
\node [draw=black, line width = 0.025cm, rounded corners = 2.5,minimum width=40, minimum height=40, text depth=0em, align=center] (user) at (5,-3.25) {Users \\[0.2em] \includegraphics[width=1cm]{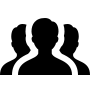}};

\draw[gray,->,>=stealth,thick] (gpu) -- node[black,above,pos=0.5,align=center,sloped,fill=white] {power $p(b)$} (n1); 
\draw[gray,->,>=stealth,transform canvas={yshift=-0.5em},thick] (user) -- node[black,below,pos=0.5,align=center] {prompts} (gpu); 
\draw[gray,->,>=stealth,transform canvas={yshift=0.5em},thick] (gpu) -- node[black,above,pos=0.5,align=center] {latency $l(b)$} (user); 
\draw[gray,->,>=stealth,thick] (user) -- node[black,above,sloped,pos=0.5,align=center] {latency measurement $\hat l$} (controller); 
\draw[gray,<-,>=stealth,shorten >= 17.5pt,thick] (controller) -- node[black,above,pos=0.8] {voltage measurement $\hat v$}  (0,1.8) -- (n1);
\draw[gray,->,>=stealth,thick] (controller.east) -- (7.75,1.8) -- (7.75,-5) -- node[black,below,pos=0.5] {batch size $b$} (0,-5)  --(gpu);
\end{tikzpicture}
}
\vspace{-2em}
\caption{GPU-to-Grid (G2G) framework for voltage regulation. The aggregated GPU behavior is simulated based on real measurement data in \cite{mlenergy-github}, and grid feedback is provided by the power system simulator (e.g., OpenDSS~\cite{opendss}). }
\label{fig:architecture}
\end{figure}
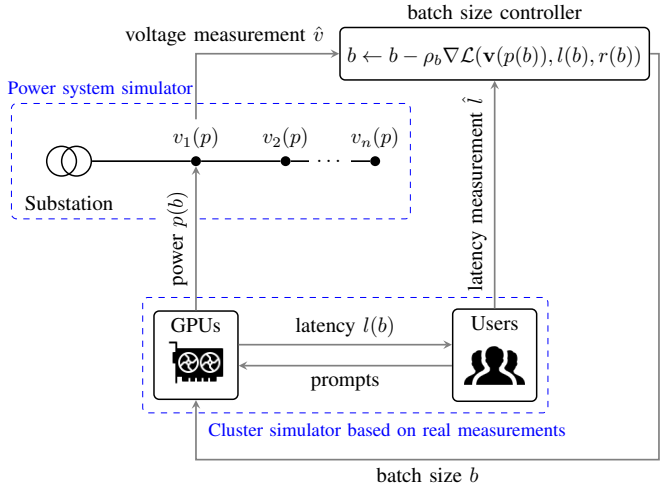

\section{Analysis of GPU Measurement Data} \label{sec:data}

\subsection{Data Source and Inference Workload Characterization}
Our work is based on real measurements from software, hardware, and workloads that are representative of modern AI data center operations.
Specifically, we used the ML.ENERGY Benchmark v3.0 data~\cite{mlenergy-neuripsdb25,mlenergy-github}, which provides detailed GPU power consumption, latency, and throughput measurements over time for various batch size\footnote{In this work, batch size refers to the LLM inference server's maximum batch size configuration, which is sustained during steady state request serving in a well-utilized datacenter.} configurations of the large language model (LLM) inference server.

Measurements were collected on vLLM~\cite{pagedattention-sosp23} v0.11.1 on NVIDIA H100 80GB SXM5 GPUs connected with NVSwitch, both of which are representative of modern AI data centers.
Workloads include dense Transformer\cite{transformer-neurips17}-based LLMs (Meta Llama 3.1 family~\cite{llama3-arxiv24}) responding to ChatGPT-style conversational queries and mixture-of-experts (MoE)~\cite{moe-arxiv17} LLMs (Qwen 3 family~\cite{qwen3-arxiv25}) answering challenging problems with reasoning, as summarized in Table~\ref{tab:llm_models}.
The models used span a variety of architectures, tasks, sizes, number of GPUs, and parameter precisions, providing substantial diversity in power consumption and performance characteristics.

\begin{table}[htbp]
\centering
\caption{LLMs studied in this paper}
\vspace{-0.8em}
\label{tab:llm_models}
\setlength{\tabcolsep}{4pt}
\begin{tabular}{llrlr}
\toprule
\textbf{Model name} & \textbf{Type} & \textbf{Active params}$^\dagger$ & \textbf{Precision} & \textbf{\#GPUs}  \\
\midrule
Llama 3.1 8B    & Dense &   8B & BF16 & 1 \\
Llama 3.1 70B   & Dense &  70B & BF16 & 4 \\
Llama 3.1 405B  & Dense & 405B & FP8  & 8 \\
Qwen3 30B A3B   & MoE   &   3B & BF16 & 2 \\
Qwen3 235B A22B & MoE   &  22B & BF16 & 8 \\
\bottomrule
\addlinespace[0.3em]
\multicolumn{5}{p{0.95\linewidth}}{\footnotesize $^\dagger$MoE models dynamically activate only a subset of total parameters.}
\end{tabular}
\end{table}

\begin{figure}[h]
    \centering
    \includegraphics[width=\linewidth]{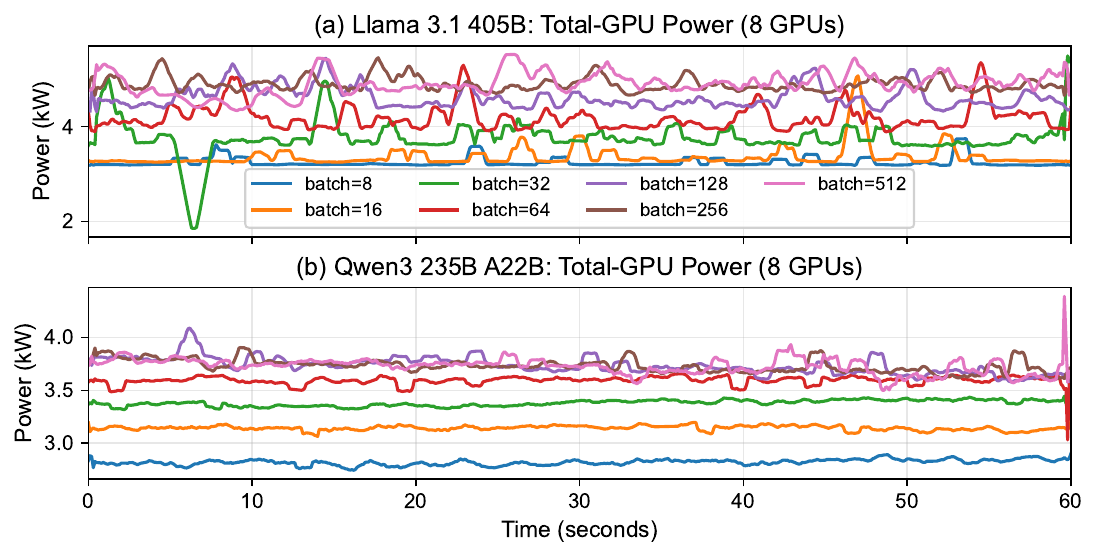}
    
    \vspace{-1.5em}

    \caption{
    Aggregated power trajectories across batch sizes for the Llama 3.1 405B and Qwen3 235B A22B models.
    }
    \label{fig:gpu_power_comparison}
\end{figure}

Fig.~\ref{fig:gpu_power_comparison} compares the GPU power consumption trajectories over time across different batch sizes for the Llama 3.1 405B and Qwen3 235B A22B models.
Both models exhibit a consistent trend: larger batch sizes result in higher average GPU power consumption.
This shared property is the foundation of batch size-based control across model architectures.



\subsection{Tradeoff between Latency and Throughput}

Inference serving performance is commonly characterized by two key metrics:
(1) latency, which quantifies the response time of individual inference requests, and
(2) throughput, which measures the total number of requests completed per unit time.
When it comes to LLMs, Inter-Token Latency (ITL) is a common latency metric, defined as the time spent to generate each output token after the previous one; a long ITL manifests as an AI chat service speaking very slowly, degrading user experience~\cite{andes-arxiv24}.
For throughput, token throughput is commonly reported, defined as the number of tokens generated per unit time; a low token throughput means that the server is not able to serve as many users at the same time, making users wait longer to get responses.

In LLM serving, tokens are generated in \emph{batches}~\cite{orca-osdi22}; $b$ requests run inference together in the GPU, and when the \emph{whole} batch has completed execution by the GPU, each request gets one new token generated (thus $b$ new tokens are generated simultaneously).
When the \emph{batch size} $b$ is increased, the raw amount of computation needed to execute inference for that batch increases.
This naturally takes more time for the GPU to complete, thereby increasing ITL.
On the other hand, with a larger batch size, the GPU's various software and hardware overheads are better amortized and the GPU's utilization increases, making it capable of completing \emph{more} computations per unit time, increasing token throughput.
A side effect of increased GPU utilization is increased power draw, as shown in Fig.~\ref{fig:gpu_power_comparison}.
%
The relationship between ITL, token throughput, and batch size for a single LLM text generation iteration can be captured by the following equation:
\begin{equation*}
{\small
\text{Token Throughput (tokens/s)}
=
\frac{\text{Batch Size (tokens)}}{\text{Inter-Token Latency (s)}}
}
\end{equation*}

%
 
Ideally, data center operators would want both high throughput and low latency.
However, these objectives are inherently in tension because increasing batch size improves throughput while simultaneously increasing latency.
We analyze and model this tradeoff relationship and the impact of batch size using real measurement data in Section~\ref{sec:metrics-and-bs}, and build our optimization model based on this relationship in Section~\ref{sec:optimization-model}.

%
%

\subsection{Relationship between Performance Metrics and Batch Size}\label{sec:metrics-and-bs}
We use GPU measurement data to empirically model the relationship between batch size $b$ and key performance metrics: (i) total GPU power consumption $p$ (in watts), (ii) mean inter-token latency $l$ (in seconds), and (iii) token throughput $r$ (in tokens per second). These relationships are represented using logistic functions, which capture the transition from underutilized to resource-saturated GPU operation as batch size increases. The logistic functions also provide inexpensive analytic gradients for online batch-size optimization.

Define the logarithmic batch size variable 
$x \triangleq \log_2(b)$.
Then the power consumption, latency, and throughput are modeled directly as functions of $x$:
\begin{align}
p(x) &= \frac{P_{\max}}{1 + \exp\!\left(-k_p (x - x_{0,p})\right)} + p_0, \label{eq:p(x)} \\
l(x) &= \frac{L_{\max}}{1 + \exp\!\left(-k_l (x - x_{0,l})\right)} + l_0, \label{eq:l(x)} \\
r(x) &= \frac{R_{\max}}{1 + \exp\!\left(-k_r (x - x_{0,r})\right)} + r_0, \label{eq:r(x)}
\end{align}
where $P_{\max}$, $L_{\max}$, and $R_{\max}$ denote the saturation magnitudes of power consumption, latency, and throughput, respectively; $k_p$, $k_l$, and $k_r$ control the sharpness of the transitions; $x_{0,p}$, $x_{0,l}$, and $x_{0,r}$ represent the characteristic batch size thresholds at which these transitions occur; and $p_0$, $l_0$, and $r_0$ are offset terms. 

The fitted relationships for the three Llama models are shown in Fig.~\ref{fig:llama_batch_fits}, while the fitting results for the two Qwen models, which exhibit similar trends, are provided in Fig.~\ref{fig:qwen_batch_fits} in Appendix~\ref{sec:appendix_A}. 
The models are fitted for the average GPU measurements over the entire observation horizon of each experiment in the ML.ENERGY benchmark dataset.

As batch size increases, GPU power consumption rises monotonically and eventually saturates. Inter-token latency also increases with batch size, with a nonlinear growth as the system approaches saturation. In contrast, token throughput initially increases rapidly with batch size, but exhibits diminishing marginal gains at larger batch sizes. These trends are consistent across model scales, although larger models operate at higher power and latency levels and reach saturation at smaller batch sizes. Overall, Fig.~\ref{fig:llama_batch_fits} demonstrates batch size is an effective control knob that induces predictable trade-offs among power, latency, and throughput with nonlinear impact.

\begin{figure}[t]
    \centering
    \includegraphics[width=\linewidth]{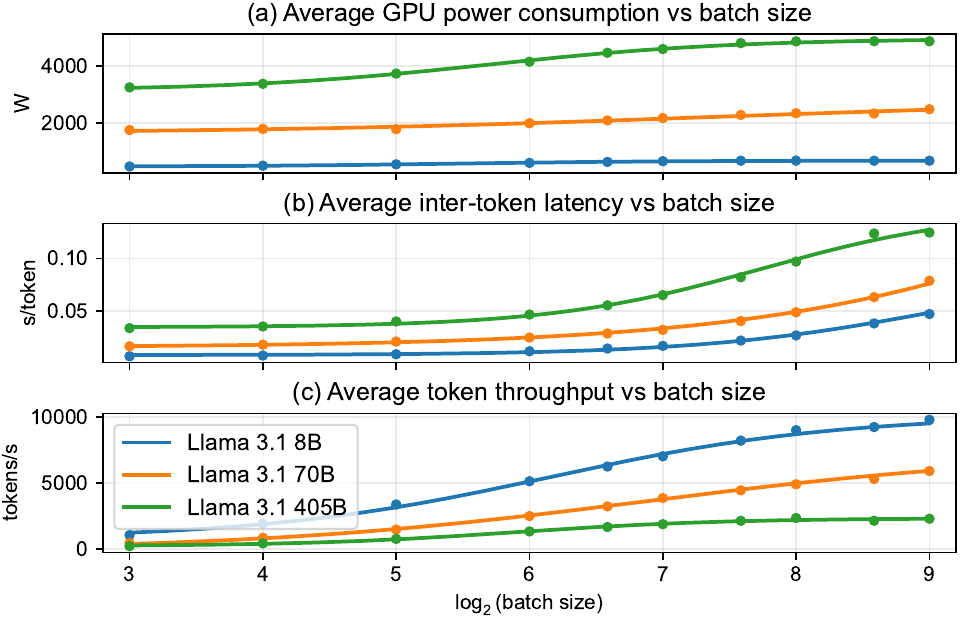}
    \vspace{-2.5em}
    \caption{
    Fitted relationships between batch size and performance metrics for three
    models in the Llama 3.1 family~\cite{llama3-arxiv24}.
    }
    
    \label{fig:llama_batch_fits}
\end{figure}

\section{Grid- and User-Aware batch size Optimization}\label{sec:optimization-model}

\subsection{Batch Size Optimization Model}
This section formulates GPU batch size control as an optimization problem, beginning with the power system model with data center load.
Consider a data center connected to a single node in a three-phase distribution network with $M$ buses. Let $\mathbf v_t \triangleq [\mathbf v_t^A,\mathbf v_t^B,\mathbf v_t^C]^\top \in \mathbb{R}^{3M}$ denote the stacked three-phase voltage magnitudes, where $\mathbf v_t^\phi\in\mathbb{R}^M$ collects voltages on phase $\phi\in\{A,B,C\}$. The phase-wise active and reactive power consumptions are $\mathbf p_t \triangleq [p_t^A,p_t^B,p_t^C]^\top$ and $\mathbf q_t \triangleq [q_t^A,q_t^B,q_t^C]^\top$, and a constant power factor $\mathrm{PF}$ is assumed for all phases, such that $q_t^\phi=\tan(\arccos(\mathrm{PF}))\,p_t^\phi$. The mapping from $\mathbf p_t$ and $\mathbf q_t$ to $\mathbf v_t$ is well established in power system modeling and simulation frameworks \cite{grainger1994power}.

Assume the data center runs inference workloads for $N$ distinct LLM models. Model $i$ is deployed using $w_i$ identical replicas, each assigned the same number of GPUs, and the vector $\mathbf w\triangleq[w_1,\dots,w_N]^\top$ collects replica counts. Under replica-based scaling, the time-averaged total power consumption and aggregate token throughput of model $i$ scale approximately linearly with $w_i$, while instantaneous power may deviate from linear scaling due to temporal misalignment of short-term fluctuations across replicas. Because replicas operate in parallel, each model's ITL does not scale with $w_i$. Accordingly, $p_i$, $r_i$, and $l_i$ denote the aggregate power, total token throughput, and average ITL of model $i$, respectively.

For simplicity, we assume that all GPUs assigned to the same model are assumed to share a common batch size configuration. Let $\mathbf b\triangleq[b_1,\dots,b_N]^\top$ denote the batch size vector, where each $b_i$ takes discrete values and is restricted to powers of two
, with batch size updates applied almost immediately after the control signals are sent to GPUs. 
To enable continuous optimization, we introduce the relaxed decision variable $\mathbf x\triangleq[x_1,\dots,x_N]^\top$, where $x_i$ approximates $\log_2(b_i)$. Since key GPU performance metrics scale smoothly in the log batch size domain, formulating the control problem in terms of $\mathbf x$ yields well-conditioned control actions.

The optimal batch size configuration can be determined by solving the following problem over $\mathbf x$ at each control interval:
\begin{subequations}\label{eq:main_opt}
\begin{align}
\max_{\mathbf x}\quad 
& \textstyle \sum_{i=1}^N r_i(x_i)
\;-\;
\gamma \, \lVert \mathbf x - \mathbf x_t \rVert_2^2
\label{eq:main_opt_obj}
\\
\text{s.t.}\quad
& \underline{\mathbf v} \;\le\; \mathbf v\!\bigl(\mathbf p(\mathbf x), \mathbf q\bigr)
\;\le\; \overline{\mathbf v}
\quad (\underline{\boldsymbol\lambda}, \overline{\boldsymbol\lambda})
\label{eq:main_opt_volt}
\\
& l_i(x_i) \le L_{\mathrm{th},i}, \quad \forall i
\quad (\mu_i)
\label{eq:main_opt_lat}
\\
& \underline x_i \le x_i \le \overline x_i, \quad \forall i ,
\label{eq:main_opt_box}
\end{align}
\end{subequations}
where $\underline x_i = \log_2(\underline b_i)$ and $\overline x_i = \log_2(\overline b_i)$ denote the lower and upper bounds on the relaxed batch size variable for model $i$.\footnote{We use $\underline b_i = 8$, as going lower hurts throughput significantly without lowering ITL. $\overline b_i$ is set as the largest batch size that fits in GPU memory.}
The optimization model's objective in \eqref{eq:main_opt_obj} aims to maximize the aggregate token throughput across all LLM models, which aligns with a fundamental operational goal of modern data centers. The regularization term $\gamma \,\lVert \mathbf x - \mathbf x_t \rVert_2^2$, with $\gamma>0$, penalizes large deviations of the current decision variable $\mathbf x$ from the previous control action $\mathbf x_t$. This term discourages abrupt changes in batch size decisions across successive control intervals, thereby promoting smoother and more stable GPU operation and corresponding power trajectories.


Moreover, the optimization model explicitly captures the coupling among three stakeholders: the power grid, the data center operator, and LLM service users. From the grid perspective, the voltage constraints in \eqref{eq:main_opt_volt} enforce three-phase voltage limits at all buses in the distribution system and are associated with dual variables $\underline{\boldsymbol\lambda}, \overline{\boldsymbol\lambda} \in \mathbb{R}^{3M}_+$. From the users' perspective, the latency constraints in \eqref{eq:main_opt_lat} impose per-model quality-of-service requirements, with dual variables $\mu_i \ge 0$. The mean inter-token latency threshold $L_{\mathrm{th},i}$ may vary across models to reflect heterogeneity in LLM architectures and service-level objectives. From the data center operator's perspective, these constraints are jointly balanced against the throughput-maximization objective, enabling batch size decisions that simultaneously respect grid reliability and user experience. 


\subsection{Batch Size Control via Online Feedback Optimization}
Since the batch size optimization in \eqref{eq:main_opt} is formulated as a continuous relaxation of an inherently integer-valued decision problem, discrepancies inevitably arise between the expected output and the realized behavior of the coupled user–GPU–grid system. Moreover, additional mismatches may be introduced by actuation delays, workload stochasticity, and unmodeled system dynamics. Online feedback optimization (OFO) inherently mitigates these issues by updating batch size decisions directly from real-time system measurements, rather than relying on exact model fidelity. This feedback-driven structure renders OFO robust to modeling inaccuracies and implementation imperfections.

We follow the standard OFO implementation in \cite{ortmann2020experimental} to solve \eqref{eq:main_opt} and add a step for discrete actuation after that. At each control interval $t = 0,1,2,\dots$, the OFO controller executes the following steps.

\paragraph*{Step 1: Measurement}
The controller measures the three-phase voltage magnitudes at all buses, denoted by $\hat{\mathbf v}_t$, and the mean ITL of each LLM model, denoted by $\hat l_{i,t}$.

\paragraph*{Step 2: Dual Variable Updates}
The dual variables associated with the voltage and latency constraints are updated via projected gradient ascent:
\begin{align}
\underline{\boldsymbol\lambda}_{t+1}
&= \Bigl[\underline{\boldsymbol\lambda}_t + \rho_v \bigl(\underline{\mathbf v} - \hat{\mathbf v}_t\bigr)\Bigr]_+,
\\
\overline{\boldsymbol\lambda}_{t+1}
&= \Bigl[\overline{\boldsymbol\lambda}_t + \rho_v \bigl(\hat{\mathbf v}_t - \overline{\mathbf v}\bigr)\Bigr]_+,
\\
\mu_{i,t+1}
&= \bigl[\mu_{i,t} + \rho_l \bigl(\hat l_{i,t} - L_{\mathrm{th},i}\bigr)\bigr]_+,
\qquad \forall i,
\end{align}
where $\rho_v > 0$ and $\rho_l > 0$ are dual step sizes, and $[\cdot]_+$ denotes element-wise projection onto the nonnegative orthant.

\paragraph*{Step 3: Primal Update in Log$_2$ Batch Size Space}
The relaxed primal decision variable $\mathbf x$ is updated via projected gradient descent:
\begin{equation}
\mathbf x_{t+1}
=
\Pi_{[\underline{\mathbf x},\overline{\mathbf x}]}
\!\left(
\mathbf x_t
-
\rho_x
\nabla_{\mathbf x}\mathcal{L}\bigl(
\mathbf x_t,
\underline{\boldsymbol\lambda}_{t+1},
\overline{\boldsymbol\lambda}_{t+1},
\boldsymbol\mu_{t+1}
\bigr)
\right), \label{eq:x_update}
\end{equation}
where $\rho_x > 0$ is the primal step size, $\mathcal{L}$ is the Lagrangian function associated with \eqref{eq:main_opt}, and $\Pi_{[\underline{\mathbf x},\overline{\mathbf x}]}(\cdot)$ denotes element-wise projection onto the box constraints $\underline x_i \le x_i \le \overline x_i$. The derivation of $\nabla_{\mathbf x}\mathcal{L}$ is provided in Appendix~\ref{sec:appendix_B}. This gradient captures the trade-offs among throughput maximization, latency constraints, voltage regulation, and penalties on large batch-size adjustments.

\paragraph*{Step 4: Discrete Actuation (Mapping $\mathbf x_{t+1}$ to $\mathbf b_{t+1}$)}
The OFO update produces a continuous decision $\mathbf x_{t+1} \in \mathbb{R}^N$, whereas the GPU runtime requires discrete batch size settings. We therefore map each component to the nearest integer in log$_2$ scale and convert back to batch size:
\begin{equation}
\tilde{x}_{i,t+1} = \operatorname{round}(x_{i,t+1}),
\qquad
b_{i,t+1} = 2^{\tilde{x}_{i,t+1}},
\qquad \forall i.
\end{equation}
The resulting batch size vector $\mathbf b_{t+1} = [b_{1,t+1}, \dots, b_{N,t+1}]^\top$ is then applied to the GPU servers.

In summary, OFO enables the data center operator to iteratively adjust GPU batch sizes using real-time voltage and latency feedback, without requiring an exact or static system model. This makes OFO particularly well suited for real-time G2G coordination under practical implementation constraints.

\section{Numerical Experiments} \label{sec:results}

\subsection{Data Center Power Profile Generation}
A key challenge in numerical studies of data centers is the lack of publicly available, high-resolution power measurements that capture responses to inference-level control knobs such as batch size. Existing datasets (e.g., the MIT Supercloud Dataset~\cite{samsi2021supercloud}) characterize aggregate behavior but do not resolve control-induced power dynamics. To address this gap, we develop a cluster simulator based on real measurement data from~\cite{mlenergy-github,mlenergy-neuripsdb25} to emulate realistic GPU responses (including power, ITL, and throughput) to batch-size control.

Synthetic load generation is designed to capture both realism and diversity. To improve realism, we superimpose multiple replica-level GPU power traces with randomly shifted start times, rather than directly scaling a single trace. This represents asynchronous workload arrivals and avoids unrealistically amplified transients. We also model time-varying inter-token latency (ITL): because historical ITL measurements exhibit heavy-tailed behavior, we fit a weighted mixture of two lognormal distributions for each batch size, as shown in Fig.~\ref{fig:llama_ITL_distribution}. At each control interval, replica-level ITLs are sampled and averaged to obtain the model-level ITL used for latency evaluation.
To introduce diversity, we include both fast and slow power variations. Fast variations are produced by a temporary training workload running concurrently with inference, representing events such as training interruptions or resumptions. Slow variations are generated by gradually reducing the number of active LLM inference replicas, mimicking changes in request arrival rates over time.

The simulated data center has an aggregate capacity of approximately 5~MW and consists of 900 servers (8 GPUs each), evenly distributed across three phases. A constant base load of 0.5~MW per phase is included to represent ancillary infrastructure such as cooling, accounting for roughly 30\% of total consumption~\cite{IEADataCenterCooling2025}. We consider five heterogeneous LLM inference workloads (detailed in Table~\ref{tab:model_settings} in Appendix~\ref{sec:appendix_C}), running together over a 60-minute horizon with 0.1~s resolution. 
A transient training workload is added over $t = 1000$ to $2000$~s, and inference demand is linearly reduced from $t=2500$ to $3000$~s, producing both short-term variability and sustained power shifts that induce significant voltage dynamics in the distribution system.
We use this workload pattern for subsequent evaluations.
Fig.~\ref{fig:dc_power_benchmark}(a) shows the resulting power profile and average ITL for a benchmark case with fixed batch size 128. Also, as shown by Fig.~\ref{fig:dc_power_benchmark}(b), the variability of per-model average ITL increases as the number of active GPUs decreases (orange area) due to reduced statistical averaging across servers.

\begin{figure}[t]
    \centering
    \includegraphics[width=\linewidth]{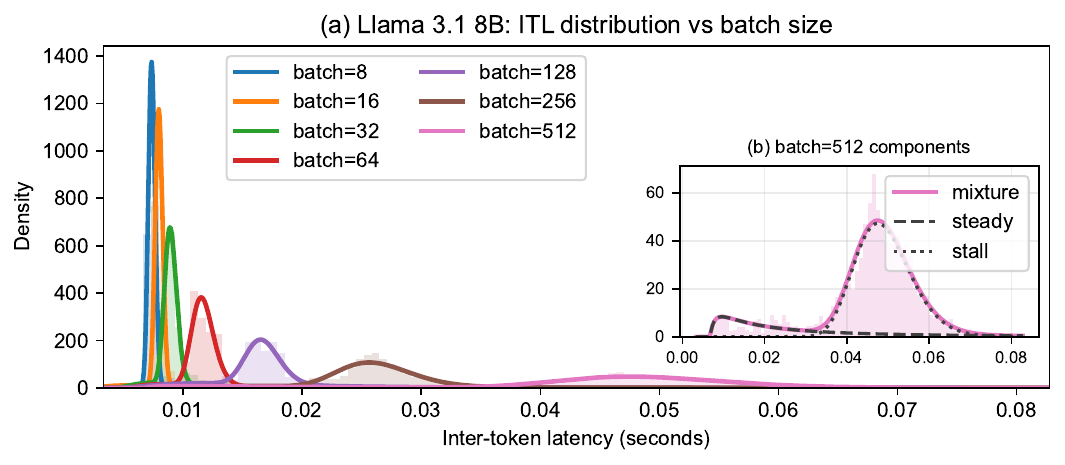}
    \vspace{-2.5em}
    \caption{Fitted ITL distributions across batch sizes for the Llama 3.1 8B model.}
\label{fig:llama_ITL_distribution}
\end{figure}

\begin{figure}[t]

    \centering
    \includegraphics[width=\linewidth]{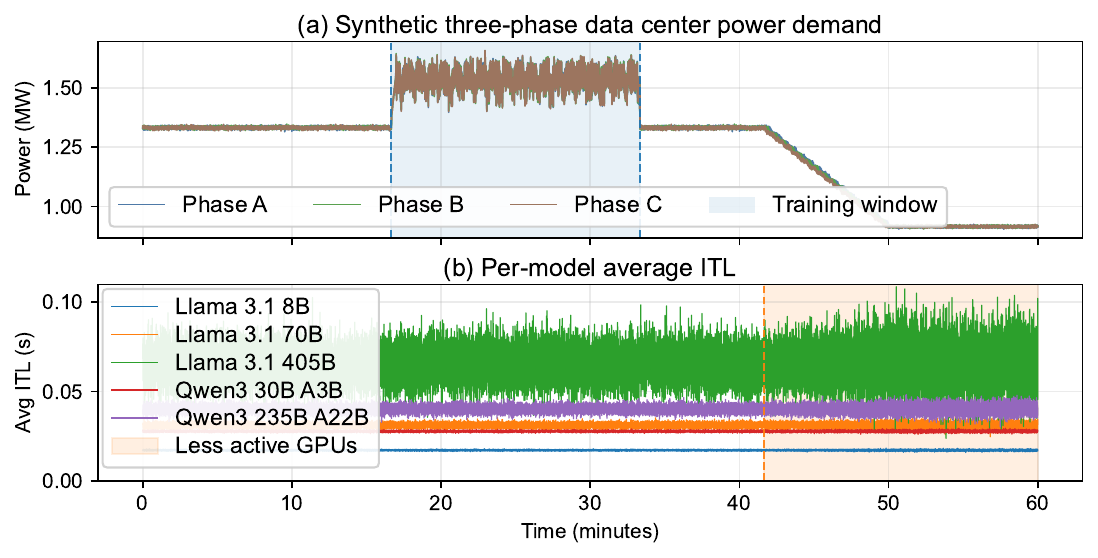}
        
    \vspace{-1.2em}
    \caption{Synthetic data center power and average ITL with a fixed batch size of 128 for all LLM models.}

\label{fig:dc_power_benchmark}
\end{figure}

\subsection{GPU-to-Grid Simulation via OpenG2G}

\begin{figure}
    \centering
    \includegraphics[width=\linewidth]{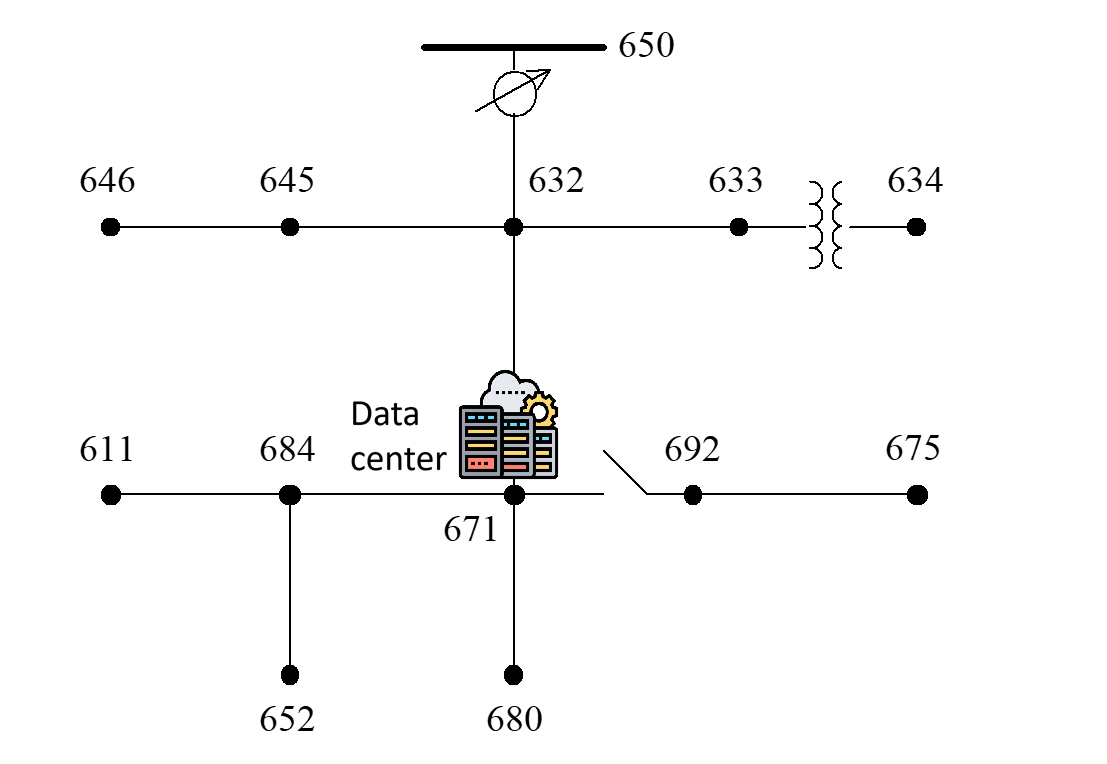}
    \vspace{-2em}
    \caption{IEEE 13-bus distribution feeder with data center load at Bus 671.}
\label{fig:IEEE_system_with_DC}
\end{figure}

We evaluate voltage impacts using the IEEE 13-bus distribution feeder~\cite{ieee13bus}, with the data center connected at Bus~671 and operating at a constant power factor of $\mathrm{PF}=0.95$, as shown in Fig.~\ref{fig:IEEE_system_with_DC}. Simulations are performed using the open-source OpenG2G library ~\cite{openg2g-arxiv26,openg2g-github}, which in turn invokes OpenDSS~\cite{opendss,opendssdirectpy} for distribution system simulation.

As a no-GPU-flexibility baseline, we simulate the synthetic data center load in Fig.~\ref{fig:dc_power_benchmark}(a), with voltage regulation provided only by step-voltage-regulator tap changes. Because frequent tap operations increase wear, maintenance needs, and outage risk, we impose a 30-minute minimum dwell time as a conservative limit on excessive mechanical actuation, with the earliest tap operation allowed at $t=25$~min. Fig.~\ref{fig:ieee13_voltages_abc_wo_ofo} shows the resulting voltage trajectories on phases~A--C. Although tap changes correct sustained deviations at $t=25$~min and $t=55$~min, the enforced delay causes temporary voltage violations after data center load changes, motivating GPU flexibility as a fast complementary voltage-regulation resource.

\subsection{Batch Size Optimization Results}

Voltage regulation with GPU flexibility is implemented using an OFO controller with primal step size $\rho_x=0.1$, dual step sizes $\rho_v=\rho_l=1$, and objective weight $\gamma=0.1$, operating at a 1~s control interval. For all models, batch sizes are selected from the discrete set $\{8,16,32,64,128,256,512\}$. The resulting voltage trajectories are shown in Fig.~\ref{fig:ieee13_voltages_abc_w_ofo}, and the corresponding GPU performance metrics are shown in Fig.~\ref{fig:ofo_simulation_results}.

To interpret the batch size trajectories in Fig.~\ref{fig:ofo_simulation_results}, we categorize controller actions into three regimes: throughput-driven, voltage-driven, and latency-driven, reflecting how the OFO controller balances data center performance objectives against grid and users' requirements.

\paragraph*{Throughput-driven regions}
When no constraints in~\eqref{eq:main_opt} are active, or when constraint violations do not dominate the gradient $\nabla_{\mathbf x}\mathcal{L}$ in \eqref{eq:x_update}, the OFO controller maximizes aggregate token throughput across all models. As shown in Fig.~\ref{fig:ofo_simulation_results}(a), throughput-driven regions appear before and after the training window, ensuring performance maximization during non-critical intervals. In our implementation, per-replica throughput is normalized to a maximum of one to enable fair aggregation across models; in practice, operators may apply model-specific throughput weights to reflect service priorities.


\paragraph*{Voltage-driven regions}
Voltage-driven actions occur during the abrupt undervoltage event near $t\approx1000$~s and the gradual overvoltage event near $t\approx3000$~s. The corresponding stepwise changes in per-replica power, highlighted in Fig.~\ref{fig:ofo_simulation_results}(b), reflect aggressive batch size reductions and increases, respectively. Comparing Fig.~\ref{fig:ieee13_voltages_abc_wo_ofo} and Fig.~\ref{fig:ieee13_voltages_abc_w_ofo}, GPU batch size control enables faster and smoother voltage recovery than tap changers, which are constrained by slow mechanical actuation. Notably, the overvoltage case illustrates that increasing GPU power consumption provides valuable grid support, a result of interest to both power and computer systems communities.

\paragraph*{Latency-driven regions}
As shown in Fig.~\ref{fig:ofo_simulation_results}(c), ITL variability increases significantly after $t\approx3000$~s. This is because empirically, larger batch sizes are associated with broader ITL distributions, resulting in greater latency fluctuations and a higher risk of violating latency constraints. Consequently, the batch size decisions in the green shaded region in Fig.~\ref{fig:ofo_simulation_results} are primarily driven by latency regulation.



 \vspace{0.5em}
Finally, Table~\ref{tab:voltage_stats_comparison} quantitatively compares the voltage regulation performance of different cases. While tap-only control prolong voltage violations relative to the uncontrolled baseline due to actuation delays and overcorrection, GPU-based control reduces the integral voltage violation (capturing both the duration and magnitude of voltage deviations) by orders of magnitude without any tap operations during the simulation. This improvement arises from closed-loop feedback, which enables rapid correction of voltage deviations and avoids the prediction errors inherent in slow, open-loop voltage regulation devices.
These results suggest that the inherent GPU flexibility may allow data centers to meet power system requirements without using additional flexible resources such as batteries.

\begin{figure}[t]
    \centering
    \includegraphics[width=\linewidth]{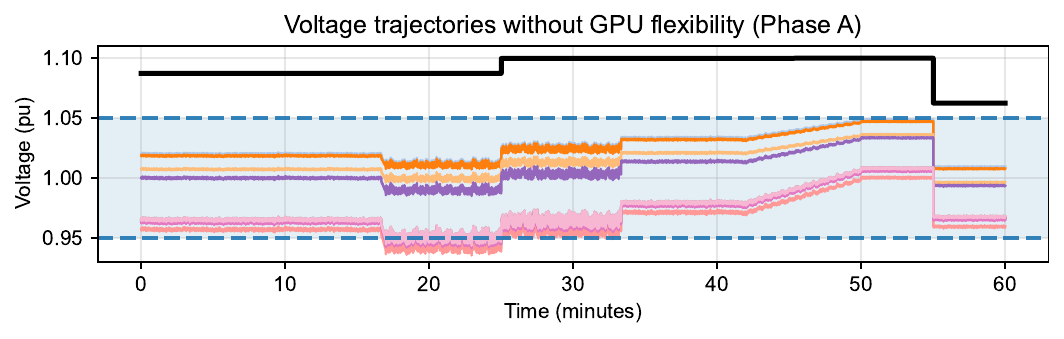}

    \vspace{-1.8em}

    \includegraphics[width=\linewidth]{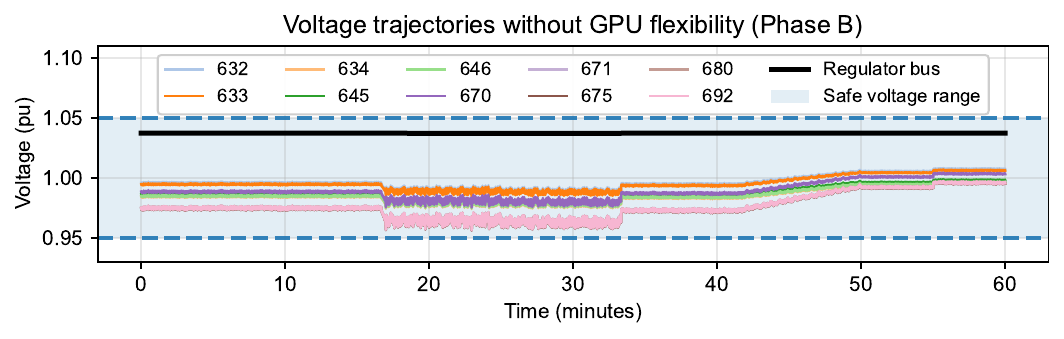}

    \vspace{-1.8em}
    \includegraphics[width=\linewidth]{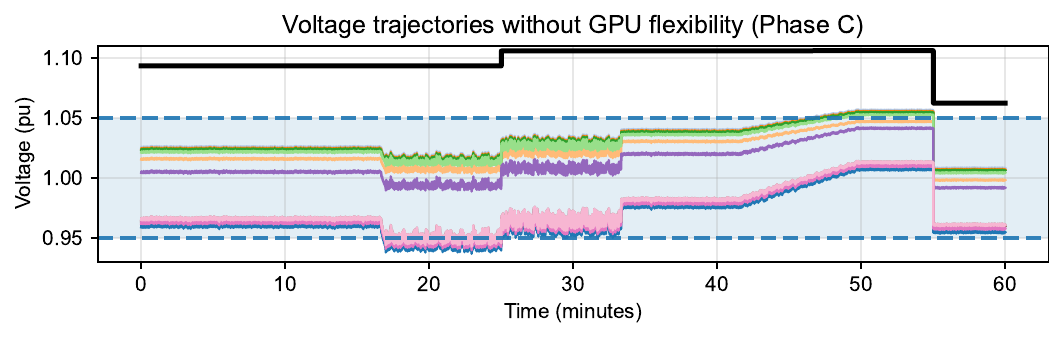}
    \vspace{-2em}

    \caption{Voltage trajectories in IEEE 13-bus system without GPU flexibility. The dashed lines indicate the voltage limits (0.95 and 1.05 pu).}
    \label{fig:ieee13_voltages_abc_wo_ofo}
\end{figure}

\begin{figure}[t]
    \centering
    \includegraphics[width=\linewidth]{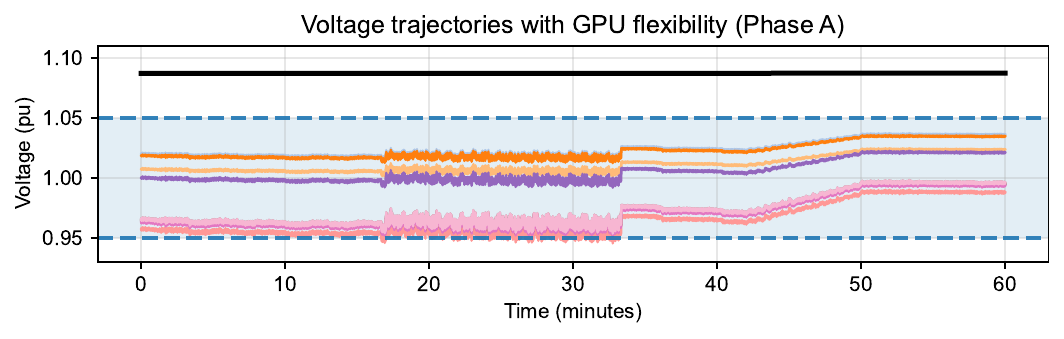}

    \vspace{-1.8em}
    \includegraphics[width=\linewidth]{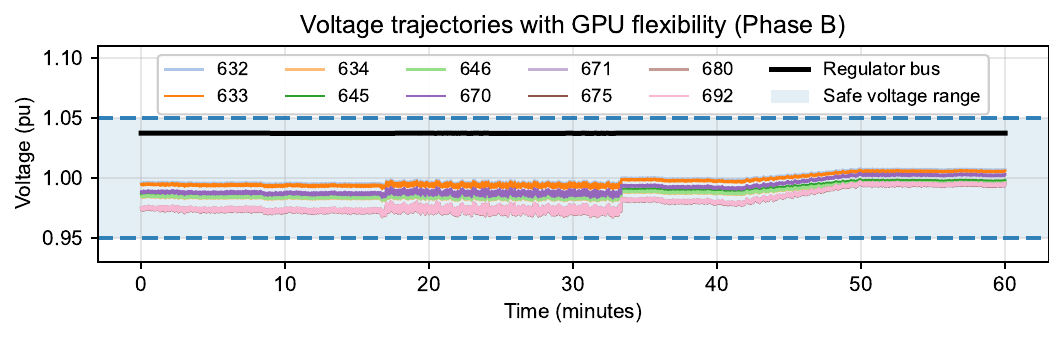}

    \vspace{-1.8em}

    \includegraphics[width=\linewidth]{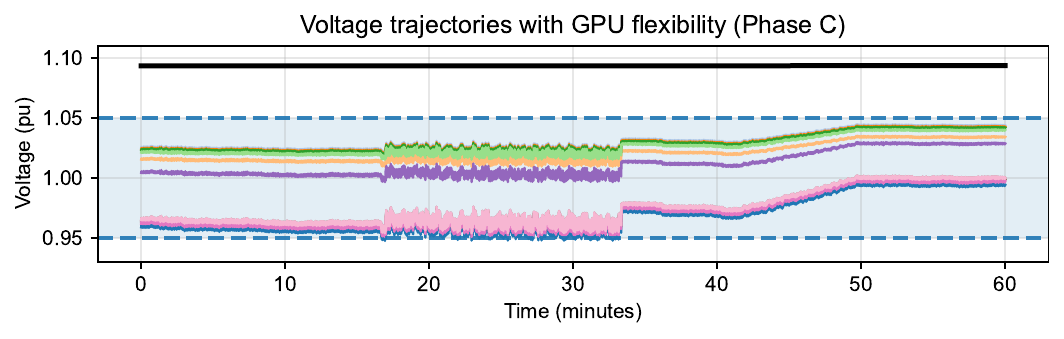}
    \vspace{-2em}
    \caption{Voltage trajectories in IEEE 13-bus system with GPU flexibility and no tap change. Voltages stay mostly within their limits.}
    \label{fig:ieee13_voltages_abc_w_ofo}
\end{figure}

\begin{figure}[t]
    \centering
    \includegraphics[width=\linewidth]{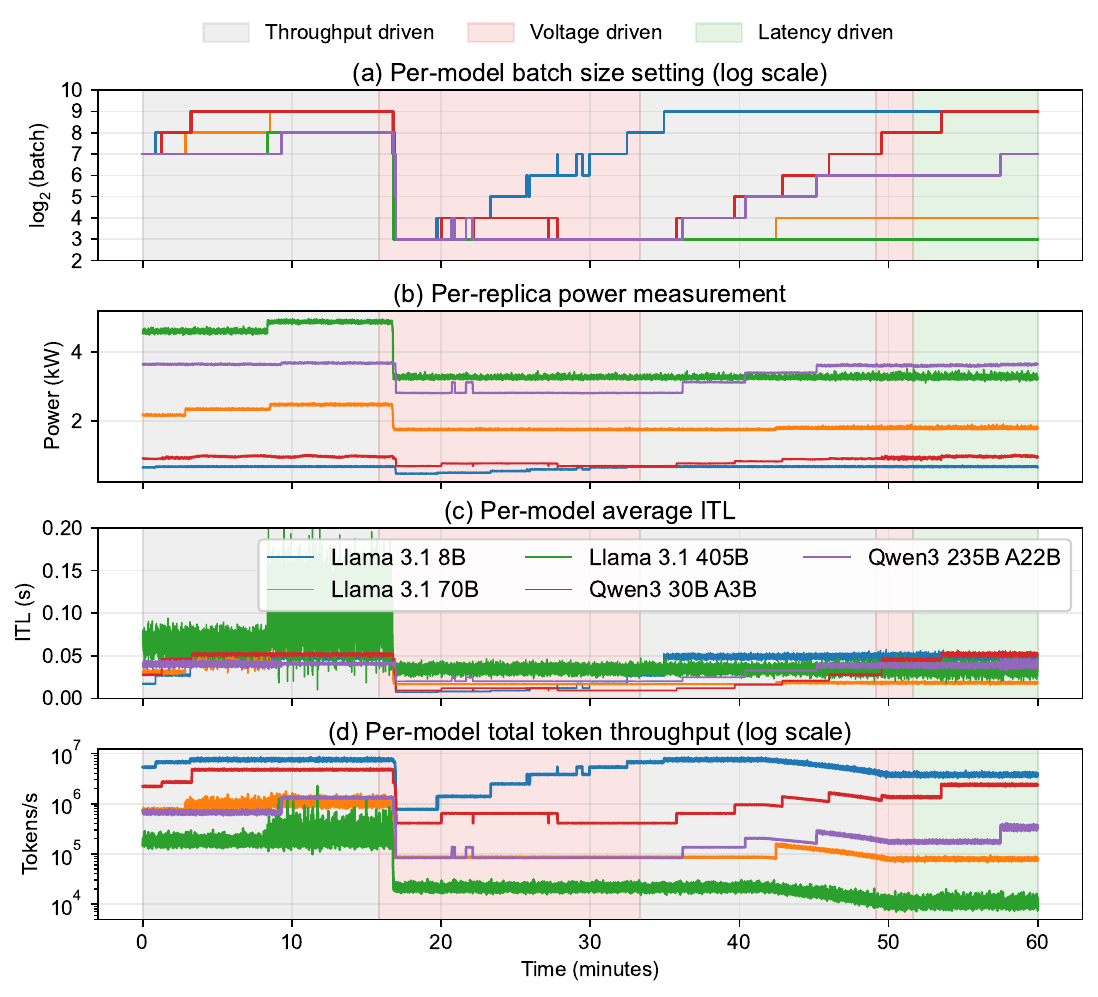}

    \vspace{-1em}

    \caption{OFO modulates batch size for each model to maximize throughput while meeting target inter-token latency constraints.}
    \label{fig:ofo_simulation_results}
\end{figure}

\begin{table}[t]
\centering
\caption{Voltage Regulation Performance Comparison}
\vspace{-0.8em}
\label{tab:voltage_stats_comparison}
\setlength{\tabcolsep}{4pt}
\begin{tabular}{lcccc}
\toprule
\textbf{Case} &
\begin{tabular}[c]{@{}c@{}}\textbf{Violation}\\\textbf{Time (s)}\end{tabular} &
\begin{tabular}[c]{@{}c@{}}\textbf{Worst}\\$V_{\min}$ \textbf{(pu)}\end{tabular} &
\begin{tabular}[c]{@{}c@{}}\textbf{Worst}\\$V_{\max}$ \textbf{(pu)}\end{tabular} &
\begin{tabular}[c]{@{}c@{}}\textbf{Integral}\\\textbf{Viol. (pu·s)}\end{tabular} \\
\midrule
No control, no tap
& 994.3
& 0.9352
& 1.0431
& 30.86 \\
Tap change only
& 1005.1
& 0.9352
& 1.0568
& 22.15 \\
GPU control only
& 62.4
& 0.9452
& 1.0445
& 0.0848\\
\bottomrule
\addlinespace[0.3em]
\multicolumn{5}{p{0.95\linewidth}}{\footnotesize Note: Violation time is total voltage violation duration. Worst $V_{\min}$/$V_{\max}$ are extrema across all buses and phases. Integral viol. is the time integral of out-of-limit voltage deviations.}

\end{tabular}
\end{table}

\section{Conclusion} \label{sec:conclusion}
This paper demonstrates the potential of GPU-level control for distribution-level voltage regulation using real LLM inference data, proving the batch size is an effective control knob for grid support by datacenters. Specifically, we propose an OFO framework that balances the requirements of the power grid, LLM service users, and data center operators by jointly considering voltage constraints, latency limits, and throughput objectives, while relying only on readily available grid measurements and avoiding the need for detailed grid information. A limitation of this study is that data center power, latency, and throughput dynamics are generated from pre-measured traces and fitted performance models, which may not capture all sources of variability present in real GPU operation. Future work includes extending to a hardware-in-the-loop setting, where GPU performance metrics are measured in real time and fully integrated into the control loop, enabling end-to-end validation under realistic operating conditions.

\section*{Acknowledgment}
We thank the reviewers for their insightful feedback.
Zhirui Liang is supported by the Eric and Wendy Schmidt AI in Science Postdoctoral Fellowship, a program of Schmidt Sciences, and Jae-Won Chung is supported by the Kwanjeong Educational Foundation and the Rackham Predoctoral Fellowship.
This work was supported in part by NSF grants CCF-2450085 and CNS-2106184, DARPA ML2P Award HR0011-26-9-E190, and grants from Ford and the Laude Institute.

The authors used AI tools to assist with narrative polishing, including spell-checking and streamlining arguments, as well as for coding and debugging. None of the narrative was directly produced by AI; it was used solely in an implementation tool and did not replace the authors. All models and ideas are the sole intellectual property of the authors, and no AI model contributed to their development.

\appendix 
\subsection{Additional GPU Measurement Data Plots} \label{sec:appendix_A}
We present additional GPU measurement results for Qwen models to demonstrate that the key findings in the main text derived from Llama models generalize to other architectures. The fitted relationships between batch size and performance metrics for two Qwen models are shown in Fig.~\ref{fig:qwen_batch_fits}, which follow the same logistic functional form as \eqref{eq:p(x)}-\eqref{eq:r(x)}. 

In addition, Fig.~\ref{fig:Qwen_ITL_distribution} shows the fitted ITL distributions across batch sizes for the Qwen3~235B~A22B model. As in Fig.~\ref{fig:llama_ITL_distribution}, each distribution is modeled as a weighted mixture of two lognormal distributions. One stall distribution represents short-duration decoding events concentrated around the mean latency, while the steady distribution captures longer-lasting components that dominate the tail of the distribution. However, for batch sizes larger than 64, the distributions exhibit greater overlap in this case, which means that the ITL is not as sensitive to batch size increase as the Llama~3.1B~8B model.

\begin{figure}[t]
    \centering
    \includegraphics[width=\linewidth]{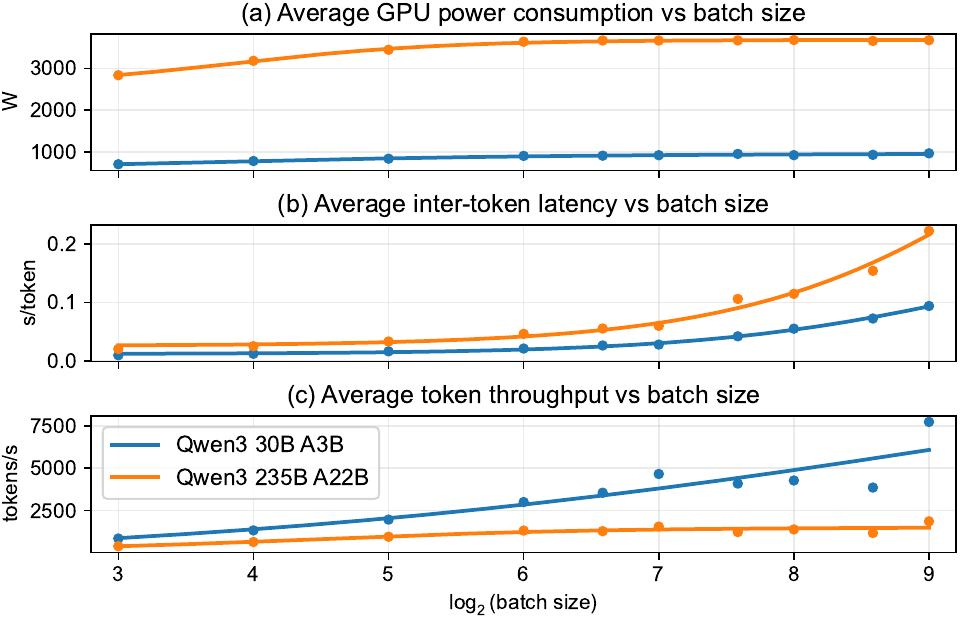}
    \vspace{-2em}
    \caption{
    Fitted relationships between batch size and performance metrics for two Qwen models.
    }
    \label{fig:qwen_batch_fits}
\end{figure}

\begin{figure}[t]
    \centering
    \includegraphics[width=\linewidth]{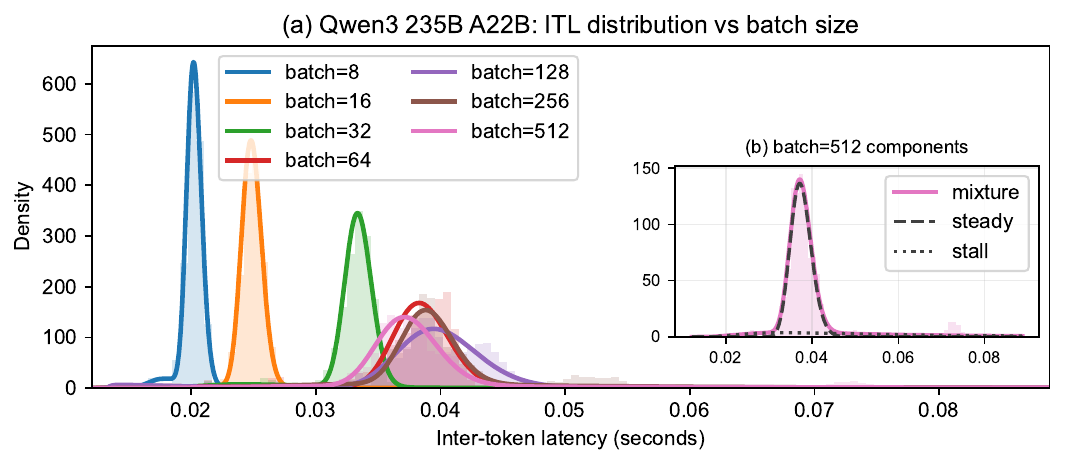}
    \vspace{-2em}
    \caption{Fitted ITL distributions across batch sizes for the Qwen3 235B A22B model.}
\label{fig:Qwen_ITL_distribution}
\end{figure}

\subsection{Gradient Derivation with Respect to Batch Size}
\label{sec:appendix_B}
The proposed formulation in \eqref{eq:main_opt} is fully differentiable with respect to $\mathbf x$. Accordingly, the associated Lagrangian function can be written as
\begin{align}
\mathcal{L}(\mathbf x,\underline{\boldsymbol\lambda},\overline{\boldsymbol\lambda},\boldsymbol\mu)
&=\;
- \textstyle \sum_{i=1}^N r_i(x_i)
+ \gamma \, \lVert \mathbf x - \mathbf x_t \rVert_2^2
\nonumber\\
&
+ \overline{\boldsymbol\lambda}^\top
\!\left[\mathbf v\!\bigl(\mathbf p(\mathbf x),\mathbf q\bigr)-\overline{\mathbf v}\right]
+ \underline{\boldsymbol\lambda}^\top
\!\left[\underline{\mathbf v}-\mathbf v\!\bigl(\mathbf p(\mathbf x),\mathbf q\bigr)\right]
\nonumber\\
&
+ \textstyle \sum_{i=1}^N \mu_i \bigl(l_i(x_i)-L_{\mathrm{th},i}\bigr).
\end{align}

Let $\boldsymbol\eta \triangleq \overline{\boldsymbol\lambda}-\underline{\boldsymbol\lambda}\in\mathbb{R}^{3M}$. The partial derivative of the Lagrangian with respect to $x_i$ is
\begin{align}
\frac{\partial \mathcal{L}}{\partial x_i}
=&\;
-\,\frac{dr_i(x_i)}{dx_i}
+ 2\gamma  (x_i - x_{t,i})
+ \mu_i \frac{dl_i(x_i)}{dx_i}
\nonumber\\
&\;
+ \boldsymbol\eta^\top \frac{\partial \mathbf v}{\partial \mathbf p}\,
\frac{\partial \mathbf p(\mathbf x)}{\partial x_i}.
\end{align}

Under a three-phase linearized distribution flow (LinDistFlow) approximation \cite{gan2014convex}, and assuming that power injections at all non–data-center buses remain constant, the bus voltage magnitudes changes from time $t$ to $t+1$ can be expressed as approximately affine functions of the power consumptions at the data center bus:
\begin{equation}
\mathbf v_{t+1} = \mathbf v_t - \mathbf R\,\Delta \mathbf p_t-\mathbf X\,\Delta \mathbf q_t,
\label{eq:lindistflow}
\end{equation}
where
$\Delta \mathbf p_t = \mathbf p_{t+1} - \mathbf p_t$
and
$\Delta \mathbf q_t = \mathbf q_{t+1} - \mathbf q_t$
denote the changes in active and reactive power consumptions at the data center bus.
The sensitivity matrices
$\mathbf R,\mathbf X\in\mathbb{R}^{3M\times 3}$
capture both within-phase and cross-phase voltage responses to variations in active and reactive data center load.
Therefore, the voltage sensitivity with respect to active power becomes
\begin{equation}
\mathbf H \triangleq \frac{\partial \mathbf v}{\partial \mathbf p}
= -\mathbf R - \tan(\arccos(\mathrm{PF}))\, \mathbf X \in\mathbb{R}^{3M \times 3}.
\end{equation}

Since model $i$ may be executed on GPUs connected to different phases $\phi \in\{A,B,C\}$ of the power system, we introduce a phase-allocation weight vector $\mathbf e_{i} = [e_{i,A}, e_{i,B}, e_{i,C}]^\top \in\mathbb{R}^3$ where $e_{i,\phi}$ denotes the fraction of GPUs assigned to model $i$ that are
connected to phase $\phi$. Therefore, we have 
\begin{equation}
\frac{\partial \mathbf p(\mathbf x)}{\partial x_i}
=
\mathbf e_{i}\,\frac{dp_i(x_i)}{dx_i},
\end{equation}

Given the logistic functions in \eqref{eq:p(x)}, \eqref{eq:l(x)}, and \eqref{eq:r(x)} which are the functions of power, latency, and throughput for one replica of model deployment, we obtain the gradient for the power, latency, and throughput of all replicas
\begin{align}
\frac{dp_i(x_i)}{dx_i}
&=
P_{\max} k_p w_i
\,
\frac{\exp\!\left(-k_p (x_i - x_{0,p})\right)}
{\left(1 + \exp\!\left(-k_p (x_i - x_{0,p})\right)\right)^2},
\\[4pt]
\frac{dl_i(x_i)}{dx_i}
&=
L_{\max} k_l
\,
\frac{\exp\!\left(-k_l (x_i - x_{0,l})\right)}
{\left(1 + \exp\!\left(-k_l (x_i - x_{0,l})\right)\right)^2},
\\[4pt]
\frac{dr_i(x_i)}{dx_i}
&=
R_{\max} k_r w_i
\,
\frac{\exp\!\left(-k_r (x_i - x_{0,r})\right)}
{\left(1 + \exp\!\left(-k_r (x_i - x_{0,r})\right)\right)^2}.
\end{align}

In summary, we obtain the gradient of Lagrangian with respect to $x_i$ as
\begin{align}
\frac{\partial \mathcal{L}}{\partial x_i}
& = \; 2\gamma  (x_i - x_{t,i})\nonumber \\
& - \; R_{\max} k_r w_i
\,
\frac{\exp\!\left(-k_r (x_i - x_{0,r})\right)}
{\left(1 + \exp\!\left(-k_r (x_i - x_{0,r})\right)\right)^2} \nonumber\\
&+\; \boldsymbol\eta^\top \mathbf H \mathbf e_{i}\,P_{\max} k_p w_i
\,
\frac{\exp\!\left(-k_p (x_i - x_{0,p})\right)}
{\left(1 + \exp\!\left(-k_p (x_i - x_{0,p})\right)\right)^2} \nonumber\\
&+\; \mu_i L_{\max} k_l
\,
\frac{\exp\!\left(-k_l (x_i - x_{0,l})\right)}
{\left(1 + \exp\!\left(-k_l (x_i - x_{0,l})\right)\right)^2}. \label{eq:final_gradient}
\end{align}

\subsection{Simulation Setup} \label{sec:appendix_C}
The topology of the IEEE 13-bus feeder with a data center load is shown in Fig.~\ref{fig:IEEE_system_with_DC}. Bus~650 serves as the upstream substation and voltage reference, with its voltage regulated by the transmission system and thus weakly influenced by downstream load variations. Voltage regulation within the feeder is primarily provided by the step-voltage regulator between Bus~650 and Bus~632, whose tap operations produce discrete voltage changes at the regulator bus in response to sustained load variations. Thus, the voltage at the regulator bus reflects discrete changes corresponding to tap operations.

In the numerical experiments, we consider the five LLM models listed in Table~\ref{tab:llm_models}. Each model is assigned an initial replica count and a latency threshold $L_{th}$, with larger models serving fewer users and tolerating higher latency. The resulting configuration occupies 600 servers, providing sufficient GPU flexibility for voltage regulation. The reset 300 servers are used for training during the training window $t\in[1000,2000]$~s.

\begin{table}[t]
\centering
\caption{LLM inference workloads and model-specific parameters in numerical experiments}
\vspace{-0.8em}
\label{tab:model_settings}
\setlength{\tabcolsep}{6pt}
\renewcommand{\arraystretch}{1.1}
\begin{tabular}{lrrr}
\toprule
\textbf{Model name} & \textbf{Replica Count} & \textbf{GPUs per replica} & \textbf{$L_{th}$ (s)} \\
\midrule
Llama 3.1 8B        & 720 & 1 & 0.08 \\
Llama 3.1 70B       & 180 & 4 & 0.10 \\
Llama 3.1 405B      & 90  & 8 & 0.12 \\
Qwen3-30B A3B      & 480 & 2 & 0.06 \\
Qwen3 235B A22B    & 210 & 8 & 0.14 \\
\bottomrule
\end{tabular}
\end{table}

\bibliographystyle{ieeetr}
\bibliography{references}

\end{document}